\documentclass[11pt]{article}

\usepackage[hyphens,spaces,obeyspaces]{url}
\usepackage{authblk}
\usepackage{float}
\usepackage{amsmath}
\usepackage{amssymb}
\usepackage{amsthm}
\usepackage[most]{tcolorbox}
\usepackage{wrapfig}
\usepackage{bold-extra}
\usepackage{lipsum}
\usepackage{amsfonts}
\usepackage{graphicx}
\usepackage{stfloats}
\usepackage{enumitem}
\usepackage{xspace}
\usepackage{url}
\usepackage{comment}
\usepackage{caption}
\usepackage{epstopdf}
\usepackage{algorithmic}
\usepackage{algorithm2e}
\usepackage{tabularx}
\usepackage{setspace}
\usepackage{todonotes}
\usepackage{xspace}
\usepackage{xcolor}
\usepackage{wrapfig}
\usepackage{array}
\usepackage{array}
\usepackage{tabulary}
\usepackage{ctable} 
\usepackage[normalem]{ulem} 
\usepackage{sober}
\usepackage{multirow,array}
\usepackage{hyperref}   
\usepackage{cite} 

\newcommand{\defn}[1]{\textbf{\emph{#1}}}

\newcommand{\POW}{PoW\xspace}
\newcommand{\genID}{\textsc{GenID}\xspace}
\newcommand{\defID}{\textsc{DefID}\xspace}

\newtheorem{problem}{Open Problem}

\date{}

\definecolor{light-gray}{gray}{0.96}

\begin{document}

\author[1]{Diksha Gupta}
\author[2]{Jared Saia\thanks{This work is supported by the National Science Foundation grants CNS-1318880 and CCF-1320994.}}
\author[3]{Maxwell Young\thanks{This work is supported by the National Science Foundation grant CCF 1613772 and by a research gift from C Spire.}}
\affil[1]{\small Dept. of Computer Science, University of New Mexico, NM, USA\hspace{6cm} \mbox{\texttt{dgupta@unm.edu}}}
\affil[2]{\small Dept. of Computer Science, University of New Mexico, NM, USA\hspace{6cm} \mbox{\texttt{saia@cs.unm.edu}}}
\affil[3]{\small Computer Science and Engineering Dept., Mississippi State University, MS, USA\hspace{6cm} \texttt{myoung@cse.msstate.edu}}

\title{Resource Burning for Permissionless Systems}

%
%
%
\maketitle              

\begin{abstract}
\vspace{5pt}
\hspace{-15pt}Proof-of-work puzzles and CAPTCHAS consume enormous amounts of energy and time.  These techniques are examples of resource burning: verifiable consumption of  resources solely to convey information.   


\hspace{5pt}Can these costs be eliminated? It seems unlikely since resource burning shares similarities with ``money burning" and ``costly signaling", which are foundational to game theory, biology, and economics.  Can these costs be reduced?  Yes, research shows we can significantly lower the asymptotic costs of resource burning in many different settings.

\hspace{5pt}In this paper, we survey the literature on resource burning; take positions based on predictions of how the tool is likely to evolve; and propose several open problems targeted at the theoretical distributed-computing research community.
\end{abstract}


\hspace{80pt}{\it ``It's not about money, it's about sending a message.''}

\hspace{267pt}{The Joker~\cite{joker}}

\section{Introduction}

In 1993, Dwork and Naor  proposed using computational puzzles to combat spam email~\cite{dwork:pricing}.  In the ensuing three decades, \defn{resource burning}---verifiable consumption of resources---has become a well-established tool in distributed security.  The resource consumed has broadened to include not just computational power, but also communication  capacity, computer memory, and human effort.

The rise of \defn{permissionless systems} has coincided with the recent increase in popularity of resource burning.  In permissionless systems, any participant---represented by a virtual identifier (\defn{ID}) in the system---is free to join and depart without scrutiny, while enjoying a high degree of anonymity.  For example, an ID might be an IP address, a digital wallet, or a username. %

In this setting, security challenges arise from the inability to link an ID to the corresponding user. A single malicious user may create a large number of accounts on a social media platform to wield greater influence; or present itself as multiple clients to disproportionately consume resources provided by the system; or inject many IDs in a peer-to-peer network to gain control over routing and content.  This malicious behavior is referred to as the \defn{Sybil attack}, originally described by Douceur~\cite{douceur02sybil}. 

Such attacks are possible because users are not ``ID-bounded'' in a permissionless system; that is, there is no cost, and therefore no limit, to the number of IDs that the attacker (\defn{adversary})  can generate. However, the adversary is often ``resource-bounded'', even if this bound is unknown.  In particular, it may be constrained, for example, in the number of machines it controls, or total channel capacity to which it has access. Resource burning leverages this constraint, forcing IDs to prove their distinct provenance by producing work that no single attacker can perform. \medskip

\noindent{\bf Paper Overview.}  Resource burning is a critical tool for defending permissionless systems.  In support of this claim, we survey an assortment of topics: distributed ledgers, application-layer distributed denial-of-service (DDoS) attacks, review spam, and secure distributed hash tables (DHTs).  Using these examples, we highlight how results in these different areas have converged upon resource burning as a critical ingredient for achieving security; this is summarized in Table~\ref{table:overview}. 
 
\begin{table}[t]
\footnotesize
  \centering
  \renewcommand{\arraystretch}{1.5}
  \begin{tabular}{>{\centering\bfseries}m{1.5cm} >{\centering}m{1.5cm} >{\centering}m{2.2cm} >{\centering}m{2.7cm} >{\centering\arraybackslash}m{2.5cm}}
    \toprule
    Domain & Primary Resource Consumed & Mechanism & Enabled Functionality  & Conjectured Cost\\
    \midrule
    Blockchains & CPU & CPU Puzzles & Distributed Ledger  & $O(\sqrt{T J_G} + J_G)$  \\
    DHTs & CPU & CPU Puzzles & Decentralized storage and search & $\tilde{O}(\sqrt{T J_G} + J_G)$ \\
    DDoS Attacks & Bandwidth / CPU & Messages / CPU Puzzles & Fair allocation of server resources &  No Conjecture\\
    Review Spam & Human Time & CAPTCHAS &  Trusted consumer recommendations &  $\tilde{O}(T^{2/3} + P_G)$ \\
    \bottomrule
  \end{tabular}
  \vspace{3pt}\caption{Summary of the domains surveyed, along with the corresponding resources, and core functionality that is secured by resource burning. We also make conjectures on the algorithmic spend rate. Here, $T$ is the adversary's spend rate; $J_G$ is the join rate for good IDs; and $P_G$ is the posting rate of good IDs.  We elaborate on these notions in Section~\ref{sec:model-main}. The $\tilde{O}$ notation omits polylogarithmic factors.} 
  \label{table:overview}
  \vspace{-15pt}
\end{table}

As prelude to this survey, we predict how resource-burning may evolve, and how systems may adapt to this technique. These predictions are distilled in four position statements below.

\vspace{-5pt}

\begin{figure}[H]
\begin{center}
\begin{tcolorbox}[standard jigsaw, opacityback=0,width=0.85\linewidth] 
\begin{minipage}[h]{1.0\textwidth}
\textbf{Position 1:} Resource burning is a fundamental tool for defending permissionless systems.
\end{minipage}
\end{tcolorbox}
\label{position-stay}
\end{center}
\vspace{-5pt}
\end{figure}

\noindent{}\POW and CAPTCHAs have been around now for decades, persisting despite concerns over scalability, resource consumption, security guarantees, and predicted obsolescence (see discussion under Position 2 and Section~\ref{sec:blockchain}). The continued practical success of resource burning aligns with theoretical justification from game-theoretic results on ``money-burning" and ``costly signaling"(Section~\ref{sec:game-theory}).  Given the increasing popularity of permissionless systems, and the need to defend them, resource burning will likely only increase in prevalence.

\vspace{-5pt}

\begin{figure}[H]
\begin{center}
\begin{tcolorbox}[standard jigsaw, opacityback=0,width=0.8\linewidth]
\begin{minipage}[h]{1.0\textwidth}
\textbf{Position 2:} Resource burning must be optimized.
\end{minipage}
\end{tcolorbox}
\label{position-burning}
\end{center}
\vspace{-5pt}
\end{figure}

\noindent In May 2020, the annual energy consumption of Bitcoin was $57.92$ terawatt-hours of electricity per year, which is comparable to the annual electricity consumption of Bangladesh; Ethereum was $7.9$ terawatt-hours, comparable to that of Angola~\cite{bc-Denmark}.  In 2012, humans spent an estimated $150,000$ hours per day solving CAPTCHAS~\cite{pogue2012time,von2008recaptcha}.  The rise of permissionless systems will likely only increase these rates of resource burning.

On the positive side, recent theoretical results suggest that resource burning can be analyzed and optimized just like any other computational resource~\cite{hartline2008optimal,Gupta_Saia_Young_2019}.  But there is significant work needed to: (1) develop a theory of resource burning focused on distributed security; and to (2) translate this theory into practical resource savings. 

In this paper, we discuss current theoretical work on reducing resource-burning rates across multiple application: blockchains (Section~\ref{sec:blockchain}); DHTs (Section~\ref{sec:dhts}); application-layer DDoS attacks (Section~\ref{sec:ddos}); and review spam (Section~\ref{sec:spam}).

\vspace{-5pt}

\begin{figure}[H]
\begin{center}
\begin{tcolorbox}[standard jigsaw, opacityback=0,width=0.8\linewidth]
\begin{minipage}[h]{1.0\textwidth}
\textbf{Position 3:} Reducing from permissionless to permissioned systems is important.
\end{minipage}
\end{tcolorbox}
\label{position-reduce}
\end{center}
\vspace{-5pt}
\end{figure}

\noindent Four decades of research have resulted in efficient and reliable algorithms for permissioned networks.  We should leverage these results when addressing problems in new permissionless systems.  One way to do this is to develop tools, based on resource burning, that bound the fraction of IDs controlled by the adversary (\defn{bad IDs}) in permissionless systems. In Sections~\ref{sec:blockchain}, we discuss results on the problem \genID, which provides this bound for static, permissionless networks; and \defID which does so for permissionless networks with churn.

In Sections~\ref{sec:ddos} and~\ref{sec:spam}, we discuss the threats posed by application-layer denial-of-service (DDoS) attacks and review spam. Neither problem aligns perfectly with a permissionless model.  For example, servers are under administrative control, and online review systems often require credentials for account creation.  However, these systems still remain vulnerable to malicious participants that are difficult to identify, and who monopolize system resources.  We define a {\it hybrid} system model as one that contains both permissioned and permissionless properties. We note that any tools designed for permissionless systems will also work for hybrid systems.  However, we would expect to be able to develop more efficient techniques to adapt tools from permissioned systems to these hybrid systems.

\vspace{-5pt}

\begin{figure}[H]
\begin{center}
\begin{tcolorbox}[standard jigsaw, opacityback=0,width=0.8\linewidth]
\begin{minipage}[h]{1.0\textwidth}
\textbf{Position 4:} Theoretical guarantees should hold independently of the resource burned. Research  should focus on both domain-specific and domain-generic problems.
\end{minipage}
\end{tcolorbox}
\label{position-evolve}
\end{center}
\vspace{-5pt}
\end{figure}

As theoreticians, we should generalize as much as possible.  Algorithms that use resource burning should require a certain ``cost" that specifies the amount of the resource to be consumed, but should allow for that resource to be anything: computation, computer memory, bandwidth, human effort, or some other resource yet to be defined.  As much as possible, theoretical results should be stated in terms of this cost, irrespective of the resource consumed.  This ensures our theoretical results will continue to be relevant, even as underlying technologies providing verifiable resource burning may change.

Additionally, a key research focus should be on problems that generalize across multiple domains.  In this paper, we describe two examples: \genID and \defID (Section~\ref{sec:gendefid}).  Our remaining three examples are domain-specific.  We believe it is important to work on both types of problems.


\section{Background and Preliminaries}

Resource burning has found application in various areas of computer security; indeed, its use was proposed by Douceur~\cite{douceur02sybil} as a defense against the Sybil attack~\cite{newsome:sybil,mohaisen:sybil,john:soft,Dinger:2006}. However, resource burning has a broader history, with similar ideas  appearing in several other scientific domains. 

In Section~\ref{sec:game-theory}, we present this background.  In Sections~\ref{sec:resourcedef},~\ref{sec:notRB}, and~\ref{sec:notwaste}, we elaborate on the notion of resource burning. Finally, in Section~\ref{sec:model-main}, we describe a general problem model that provides a unifying set of assumptions and terminology used throughout this document.

\subsection{Game Theory, Biology and Economics}\label{sec:game-theory}

Resource burning is analogous to what is referred to as \emph{money burning} in the game theory literature.  To the best of our knowledge, the first significant algorithmic game theory study of money burning, due to Hartline and Roughgarden, analyzed the use of money burning in mechanism design~\cite{hartline2008optimal}.  Their main result is a near-optimal mechanism for multi-unit auctions, where the quantity optimized is \defn{social welfare} or the sum of utilities of all players.  They also give results showing that, under certain conditions, an auction utilizing money burning can obtain a $\Omega\left(\frac{1}{1+ \log(n/k)}\right)$ fraction of the optimal social welfare, where the auction consists of $n$ bidders who are bidding for $k$ units.  They conclude that \emph{``the cost of implementing money-burning ... is relatively modest, provided an optimal money-burning mechanism is used"}.

Money burning is also known as \emph{costly signaling} in the game theory literature, and it has two main uses in this context.  First, it can signal commitment to a certain action, as is illustrated in the ``lunch" game\footnote{This is equivalent to what is referred to as the ``battle of the sexes" game in~\cite{binmore2007playing}}~\cite{huck2005burning,binmore2007playing}.  Second, it can signal the ``type" of a player, as is in the``college" game~\cite{binmore2007playing}. We present these two games below.

\medskip
\noindent
\textbf{Lunch Game.} Two friends want to eat lunch together, but the first friend prefers option A and the second prefers option B.  They each  obtain payoff of $-1$ if they choose different locations.  If they both pick option A, they  obtain payoffs of $10$ and $1$ respectively.  Conversely, if they both pick option B, they obtain payoffs of $1$ and $10$.  

Now, if the first friend verifiably burns money equal to $1$ unit of utility prior to playing the game, this signals a commitment to their preferred option, since if they were to choose the unpreferable option, their utility would now be at most $0$.  Thus, they would not have played the game.  In this way, a friend who burns money can expect higher utility.

\medskip
\noindent
\textbf{College Game.}  Each student is one of two types: \emph{smart} or \emph{daft}.  Each student is considering college and can choose either the action \emph{attend} or \emph{not attend}.  A smart student pays a cost of $1$ (in terms of time and effort) to attend college, and a daft student pays a cost of $3$ to attend college.  We assume that the decision of the student to attend college is publicly known, but that otherwise, college has no impact: daft students stay daft even after attending.\footnote{On the positive side, smart students stay smart!}

An employer wants to hire smart students.  If the employer hires a smart student, their benefit is $2$, and if they hire a daft student, their cost is $2$.  If a student is hired by the employer, they have a benefit of $2$, and if they are not hired, they have a benefit of $0$.

It is easy to verify that the following is a Nash equilibrium for this game:
\begin{itemize}\renewcommand{\labelitemi}{$\circ$}
	\item Smart students attend college.
	\item Daft students do not attend college.
	\item The employer hires only students that attend college.
\end{itemize}

Here, smart students all choose to attend, \emph{even though college has no intrinsic benefit}.  Thus, the choice to attend college is a costly signal made by the smart students, and college itself is an example of resource burning.

If the option to attend college were removed from the game, and the fraction of smart students were less than $1/2$, then a Nash equilibrium would be for the employer to never hire.  In this case, the overall social welfare---the sum of expected benefits to all players---would decrease.

\medskip
\noindent
\textbf{Biology.} Costly signaling is a well-known phenomena in biology.  A relevant example from animal behavior is \emph{stotting}, in which quadrupeds, such as deer and gazelles, repeatedly jump high into the air.  This is often done in view of a predator, suggesting that stotting is a costly signal to the predator that the prey is too healthy to catch~\cite{fitzgibbon1988stotting}.  Other examples occur in sexual-selection, where the use of plumage, large antlers, and loud cries are a costly signal of fitness~\cite{zahavi1975mate}.

\medskip
\noindent
\textbf{Economics.} In 1912, the economist Thostein Veblen coined the term ``conspicuous consumption" to describe costly signaling used by people to advertise both wealth and leisure.  For example, Veblen writes, \emph{``The walking stick serves the purpose of an advertisement that the bearer's hands are employed otherwise than in useful effort, and it therefore has utility as an evidence of leisure"}~\cite{thorstein1912theory}.  Decades of economic studies suggest that conspicuous consumption is a critical part of historical and modern economies~\cite{miller2009spent,penn2003evolutionary,saad2007evolutionary,saad2009effect,sundie2011peacocks}.  For example, Sundie et al write: \emph{``Although showy spending is often perceived as wasteful, frivolous, and even narcissistic, an evolutionary perspective suggests that blatant displays of resources may serve an important function, namely, as a communication strategy designed to gain reproductive reward"}\cite{sundie2011peacocks}.


\subsection{What is Resource Burning?}\label{sec:resourcedef}

We define \defn{resource burning} as the verifiable consumption of a resource.  In particular, it is computationally easy to verify both the consumption of the resource, and also the ID that consumed the resource~\cite{ali:foundations}.  Below we describe several resource-burning techniques. 

\smallskip
\noindent{\bf Proof-of-work (\POW).} \POW is arguably the current, best-known example of resource burning. Here, the resource is computational power.  Proof-of-work has been proposed for spam-prevention~\cite{dwork:pricing,liu:proof,laurie-proof}; blockchains~\cite{nakamoto:bitcoin}; and defense against Sybil attacks~\cite{li:sybilcontrol,andrychowicz2015pow}.

\smallskip
\noindent{\bf CAPTCHAs.} A {\it completely automated public Turing test to tell computers and humans apart}, or a CAPTCHA, is a resource-burning tool where the resource is human effort~\cite{5601666}.  CAPTCHAs may be based around text, images, or audio; however,  several design and usability issues exist~\cite{yan:usability}.

\smallskip
\noindent{\bf Proof-of-Space.} Proof-of-space requires a prover to demonstrate utilization of a certain amount of storage space~\cite{dziembowski2015proofs,ateniese2014proofs,abadi2005moderately,dwork2003memory}. This approach is foundational for Spacemint cryptocurrency~\cite{park2018spacemint}.  Like \POW, proof of space demonstrates the consumption of a certain amount of a physical resource, but can require less electrical power.  A related proposal is ``Proof of Space-Time"~\cite{moran2019simple}, which demands proof of consumption of a certain amount of storage space for a certain amount of time.
 
\smallskip
\noindent{\bf Resource Testing.} Resource testing requires a prover to demonstrate  utilization of a radio channel~\cite{gilbert:sybilcast,gilbert:who,monica:radio}.\footnote{Resource burning refers to the game-theoretic money burning technique; resource testing  refers to that technique specifically applied in the wireless domain.}  Consider a wireless setting where each device has a single radio that provides access to one of several channels. Thus, an adversary representing two bad IDs, but with a single device, can only listen to one channel at a time. A base station can assign each ID to separate channels; send a random message on one of these channels chosen randomly; and demand that the message be echoed back by the corresponding ID. Since the adversary can only listen to a one channel at a time, it will fail this test with probability at least $1/2$.  \smallskip


\subsection{What is \emph{not} Resource Burning} \label{sec:notRB}

\defn{Proof-of-Stake (PoS)} is a defense for permissionless systems, wherein security relies on the adversary holding a minority stake in an abstract finite resource~\cite{abraham:blockchain}.  It has been proposed primarily for cryptocurrency systems (Section~\ref{sec:blockchain}).  When making a group decision, PoS ensures that each ID has voting weight proportional to the amount of cryptocurrency that ID holds.  Well-known examples of such systems are ALGORAND~\cite{GiladHMVZ17}, which employs PoS to form a committee, and Ouroboros~\cite{Kiayias2017}, which elects leaders with probability proportional to their stake.  Hybrid approaches using both \POW and PoS exist, including one proposed for the Ethereum system~\cite{ethereum-pos}, and under the name ``Proof of Activity"~\cite{bentov2014proof}.  In contrast to the above examples,  PoS involves a measurement, rather than a consumption of, a resource.  

\smallskip
\noindent{\bf Disadvantages of Proof-of-Stake.} 
Unfortunately, PoS can only be used in systems where the ``stake" of each ID is globally known.  Thus, it seems likely to remain relevant primarily in the  domain of cryptocurrencies.  Moreover, even within that community, there are concerns about proof-of-stake.  To quote  researcher Dahlia Malkhi: \emph{``I think proof-of-stake is fundamentally vulnerable \ldots In my opinion, it's giving power to people who have lots of money"}~\cite{posdm}.

\subsection{Resource Burning Does Not Require \emph{Waste} of the Resource}\label{sec:notwaste}

While resource burning requires verifiable \emph{consumption} of a resource, it does not necessarily require \emph{waste} of that resource.  For example, Von Ahn et al.~\cite{von2008recaptcha} developed the reCAPTCHA system which channeled human effort from solving CAPTCHAs into the problem of deciphering scanned words that could not be recognized by computer.  Their system achieved an accuracy exceeding professional human transcribers, and was responsible for sucesssfully transcribing hundreds of millions of words from public domain books.

In 2018, Ball et al. developed proof-of-work puzzles whose hardness is based on worst-case assumptions~\cite{ball2018proofs}.  These puzzles are based on the Orthogonal Vectors, 3SUM, and All-Pairs Shortest Path problems, and any problem that reduces to these problems, including deciding any graph property statable in first-order logic.  Hence, their work enables design of \POW puzzles that can be useful for solving computational problems of practical importance.  

In~\cite{shoker2017sustainable}, Shoker developed proof-of-work puzzles that solve real-world matrix-based scientific computation problems.  He named this technique ``Proof of Exercise".

All algorithms discussed in this paper are compatible with this type of ``useful" resource burning, where the consumption of the resource solves practical problems.  Our only requirement of the resource burning mechanism is that the consumption of the resource be easily verifiable, which holds true for the above results.

\subsection{A General Model}\label{sec:model-main}

We discuss broad aspects of a general model for permissionless systems. This allows us to highlight commonalities between different application domains, while retaining the same terminology throughout.  

The system consists of virtual \defn{identifiers (IDs)}. An  ID is  \defn{good} if it obeys protocol and belongs to a unique user; otherwise, the ID is \defn{bad}.  Good and bad IDs cannot necessarily be distinguished {\it a priori}.\smallskip

\noindent{\bf Communication.}\label{sec:com} Communication is synchronous and occurs either via point-to-point or via a broadcast primitive. The former is typical for peer-to-peer systems and the general client-server setting.  The latter corresponds to permissionless blockchains, where it is a standard assumption that  a  good  ID  may  send  a  value  to all other good IDs within a known and bounded amount of time, despite an adversary; for examples, see \cite{Garay2015,bitcoinwiki,GiladHMVZ17,Luu:2016} and see~\cite{miller:discovering} for empirical justification. 

\smallskip

\noindent{\bf Adversary.} A single adversary controls all bad IDs; this pessimistically represents perfect collusion and coordination by malicious users. Bad IDs may arbitrarily deviate from our protocol, including sending incorrect or spurious messages. The adversary can send messages to any ID at will, and can view any communications sent by good IDs before sending its own.  It knows when good IDs join and depart, but it does not know in advance the private random bits generated by any good ID.   

Often, the adversary is assumed to control only an $\alpha$-fraction of the network resources, for $\alpha>0$. Generally,  in settings where correctness is threatened, $\alpha$ must be a small constant; for example, often bounded below $1/3$ or $1/4$.  Alternatively, there are settings where $\alpha$ can be any constant bounded away from $1$; typically, this corresponds to  problems of performance (rather than correctness).

\smallskip
\noindent {\bf Tunable Costs.} We measure \defn{cost} as the amount of resource consumed.  Our model is agnostic with respect to the particular resource used.  However, we assume that it is possible to arbitrarily tune the cost.  In particular, we assume that, for any value $x$, an ID can be issued a \defn{challenge} of difficulty $x$ that will require consumption of $x$ units of whatever resource is used.

Resources such as computation, computer memory, and bandwidth have inherently tunable costs.  For CAPTCHAs, cost could be adjusted in two possible ways.  First, by adjusting the difficulty of the puzzle, by either (1) adjusting the number of alphanumeric digits or the number of images to be classified; or (2) adjusting the difficulty of an individual recognition task as described in the ScatterType CAPTCHA system~\cite{baird2005scattertype}.  Second, by adjusting the \emph{expected} difficulty by adjusting a probability of being required to solve a CAPTCHA.

\smallskip
\noindent{\bf Joins and Departures.}\label{sec:join} Often, the system is dynamic, with IDs joining or departing over time. There is no {\it a priori} method for determining whether a joining ID is good or bad. Joins and departures by bad IDs may be scheduled in a worst-case fashion, and pessimistically we often assume the adversary also has a limited ability to schedule these events for good IDs.  We will generally assume a lower bound on the number of IDs in the system, and that the lifetime of the system is polynomial in this lower bound.

\smallskip
\noindent {\bf Key Notation.}
Through out this work, let {\boldmath{$T$}} denote the  \defn{adversarial spending rate}, which is the cost to the adversary over the system lifetime divided by the lifetime of the system.  Let the \defn{algorithmic spending rate},~{\boldmath{$A$}}, be the cost to all good IDs over the system lifetime divided by the lifetime of the system.  

In the blockchain and DHT problems, we let {\boldmath{$J_G$}} denote the \defn{good ID join rate}, which is the number of good IDs that join during the system lifetime divided by the lifetime of the system.  Finally, for the review spam problem, we let {\boldmath{$P_G$}} denote the \defn{good posting rate}, which is the number of posts made by good IDs during the system lifetime divided  by lifetime of the system.

\subsection{Game Theoretic Analysis}

For many of our problems, we can analyze the defense of a system as a two-player zero sum game~\cite{easley2010networks} as follows.  There is an adversary that can choose to attack or not, and an algorithm that can choose to defend or not.  There is a  system invariant, which the algorithm seeks to protect, that has some value $V$.  
There is a function $f$ that gives the cost incurred when the algorithm chooses to defend as follows: if the adversary spends $T$ to attack, then the algorithm will spend $f(T)$ to defend. Thus the payoff matrix for the algorithm is given below. 

\vspace{-7pt}

\begin{table}[H]
    \setlength{\extrarowheight}{2pt}
    \begin{tabular}{p{3cm}cc|c|c|}
      & & \multicolumn{1}{c}{} & \multicolumn{2}{c}{Adversary}\\
      & & \multicolumn{1}{c}{} & \multicolumn{1}{c}{Attack}  & \multicolumn{1}{c}{$\neg$Attack} \\\cline{4-5}
      \multirow{2}*{ \hspace{2cm}Algorithm}  & & Defend & $T-f(T)$ & $-f(0)$ \\\cline{4-5}
      & & $\neg$Defend & $-V$ & $0$ \\\cline{4-5}
    \end{tabular}
\end{table}

\vspace{-5pt}

Solving this game, we get that in the Nash equilibrium, the algorithm player will defend with probability $p = \frac{V}{T-f(T) + f(0) + V}$.  Thus, the expected utility of the game to the algorithm player will be $\frac{-Vf(0)}{T-f(T) + f(0) + V}$.  In many of our problems, $f(T) = f(0) + o(T)$, and so we obtain a value that is $\Theta\left(\frac{-Vf(0)}{T + V}\right)$.  Smaller $T$ optimizes the utility for the adversary, in which case, the expected utility of the algorithm is $\Theta(-f(0))$.

\section{Blockchains and Cryptocurrencies}\label{sec:blockchain}

A \emph{blockchain} is a distributed ledger.  In particular, it is a distributed data structure that stores transactions between IDs in a network.  Each transaction represents flow of a resource from one ID to another. Every transaction added must be legitimate, in the sense that the source ID owns the resource to be transferred, as indicated by the distributed ledger, at the time of the transaction.  Importantly, transactions can only be added to the blockchain, and once added, can never be deleted or edited.  



\subsection{\genID and \defID}\label{sec:gendefid} 

Perhaps the current, most frequently-used application of resource-burning is for blockchains.  Permissionless blockchains are vulnerable to Sybil attacks~\cite{lin2017survey}.  The next two problems use resource burning to defend against this.  Recall that the adversary controls an $\alpha$-fraction of the resource that is being burned. \smallskip

\noindent{\bf The \genID Problem.} The problem stated below, \genID, was first defined and studied by Aspnes, Jackson, and Krishnamurthy~\cite{AspnesJK2005}. They proposed a solution with latency of $3$ rounds, and $\tilde{O}(n^2)$ bits sent per good ID, at a burned resource cost of $O(1)$ per good ID. 

\vspace{-15pt}

\begin{figure}[H]
\centering
\begin{tcolorbox}[standard jigsaw, opacityback=0]
\begin{minipage}[h]{1.0\textwidth}

\begin{problem}\label{op:genID}
\noindent{\bf \genID}
\end{problem}

\noindent{\bf Model:} Initial set of IDs; $n$ of which are good, with the rest are controlled by an adversary.

\smallskip

\noindent{\bf Goal:} All good IDs decide on a set of IDs $S$ such that: (1) all good IDs are in $S$; and (2) at most a $O(\alpha)$ fraction of the IDs in $S$ are adversarial. 
\end{minipage}
\end{tcolorbox}
\label{position-gen}
\end{figure}

\vspace{-5pt}

\noindent Several other solutions to \genID have been proposed in the literature~\cite{andrychowicz2015pow,hou2017randomized,aggarwal2019bootstrapping,katz2014pseudonymous}. Andrychowicz and Dziembowski described an algorithm  with a latency of $\Theta(n)$ rounds; $\tilde O(n^2)$ bits sent per good ID; and a burned resource cost of $\tilde{O}(1)$ per good ID~\cite{andrychowicz2015pow}.   Concurrent to this work, Katz, Miller and Shi \cite{katz2014pseudonymous} proposed another solution with similar costs.  Hao et al.~\cite{hou2017randomized} improved on these results via using a randomized leader election protocol. Their algorithm has, in expectation, a latency of $\Theta\left(\frac{\ln n}{\ln \ln n}\right)$ rounds; $\tilde{O}(n)$ bits sent per good ID; and a burned resource cost of $\Theta\left(\frac{\ln n}{\ln \ln n}\right)$ per good ID.  

The most recent work in this domain is by Aggarwal et al.~\cite{aggarwal2019bootstrapping}, which requires in expectation: $O(1)$ latency; $O(n)$ bits sent per good ID; and a burned resource cost of $O(1)$  per good ID.  

It is still not known if these costs can be reduced for the general problem, or for an ``almost-everywhere" versions of the problem, where all but a $o(1)$ fraction of the IDs must learn $S$.  To the best of our knowledge, there are no current lower-bounds on the problem.

\smallskip

\noindent\textbf{The \defID Problem.} The following problem, called \defID, considers the \genID problem in the presence of churn.

\vspace{-5pt}

\begin{figure}[H]
\centering
\begin{tcolorbox}[standard jigsaw, opacityback=0]
\begin{minipage}[h]{1.0\textwidth}

\begin{problem}\label{op:defendID}
\noindent{\bf\textsc{DefID}}
\end{problem}

\noindent{\bf Model:} Stream of IDs joining and leaving a network.

\smallskip

\noindent{\bf Goal:} At most an $O(\alpha)$-fraction of bad IDs in the network at any time. 

\end{minipage}
\end{tcolorbox}
\label{position-def}
\end{figure}

\vspace{-5pt}

A first algorithm to solve \textsc{DefID} was proposed in by Gupta, Saia and Young in \cite{pow-without}.  It required algorithmic spend rate of $O(J_G + T)$; recall that $J_G$ is the join rate of good IDs per time step, and $T$ is the spend rate of the adversary. Note that this result holds without any additional assumptions. Gupta, Saia and Young further improved this result in \cite{Gupta_Saia_Young_2019,gupta2020proof} to $O(J_G + \sqrt{TJ_G})$, subject to two assumptions on the join rate of good IDs, which are found to be supported by real-world data\cite{Gupta_Saia_Young_2019}.
 
 Specifically, the assumptions needed are as follows.  Define an \defn{epoch} to be the length of time it takes for the fraction of good IDs to change by 3/4 fraction.  First, the join rate for good IDs changes by at most a multiplicative factor between any two successive epochs. Second, in any epoch the actual join rate for good IDs over any ``sufficiently large" period of time is within constant factors of the join rate for good IDs over the entire epoch.
 
An asymptotically matching lower bound was obtained for a large class of algorithms~\cite{Gupta_Saia_Young_2019}. An open problem is to generalize this bound to all algorithms.


\section{Distributed Hash Tables}\label{sec:dhts}

Distributed hash tables (DHTs) are a popular P2P distributed data structure~\cite{stoica_etal:chord,kashoek_karger:koorde,li:fissione,abraham:land,maymounkov:kademlia,rowstron_druschel:pastry} with several implementations over the years~\cite{6688697,steiner:global,falkner:profiling}. Generally, the design entails hashing attributes of a user's machine to a key value (or ID) in a virtual space; similarly, for data items. The various DHT constructions differ in their overlay topologies, but typically IDs need only maintain state on a small number of neighbors, and routing is possible with a small number of messages, where small means at most logarithmic in the number of IDs in the system.

These systems are vulnerable to attack. A bad ID that participates in routing can drop or corrupt any message it receives. A good ID can be completely isolated from the rest of the network if all of its neighbors are bad; this is often referred to as an \defn{eclipse attack}~\cite{190890,4146884}. Finally, content can be compromised if bad IDs alone are responsible for storing a particular data item.  Generally, the behavior of bad IDs is modeled by Byzantine faults.  For almost two decades, there has been a sustained interest in the design of secure DHTs that can tolerate such attacks~\cite{urdaneta:survey}.\medskip

\noindent{\bf Byzantine Fault Tolerance in DHTs.}  A popular approach to tolerating bad IDs depends makes  use of  \defn{groups}: these are small sets of IDs, each of which have a good majority. Intuitively, a group is used in place of an individual peer, and the group members  act by using majority action or secure multiparty computation to coordinate actions. For example, routing can be performed robustly via all-to-all communication between each pair of groups along the path from source to destination. Examples of group-based DHT constructions include~\cite{fiat:making,awerbuch:towards,awerbuch:random,awerbuch:towards2,saia:reducing,young:towards,naor:novel,JaiyeolaPSYZ18,sen:commensal}.

As an alternative to using groups,  bad IDs may be tolerated by  employing some form of redundant routing~\cite{halo:kapadia,salsa:nambiar,salsa2:safwan,castro:secure,HK,johansen:fireflies}. Several other results do not explicitly apply to DHTs, although they may be compatible. For example, the challenge of tolerating bad IDs is exacerbated in highly-dynamic P2P systems, and there is a growing body of work in this area~\cite{guerraoui:highly,augustine:fast,Augustine2015,7354403,Augustine:2012,Augustine:2013:SSD:2486159.2486170}. Self-healing networks are another approach for achieving security, where bad IDs are identified and evicted~\cite{saad:self-healing,saad2017theoretical,saad:self-healing2}.\smallskip

In all of these works, a critical assumption is that the fraction of bad IDs is a small constant. However, given that DHTs are often permissionless, this assumption is easily violated via a Sybil attack. Thus, while many tools have been developed for securing DHTs against Byzantine faults, additional work is required to limit the fraction of bad IDs in the permissionless setting.\smallskip

\noindent{\bf Sybil Resistance.} Several approaches have been proposed for mitigating the Sybil attack. The influence of bad IDs can be limited via containment schemes that leverage the network topology in structured overlays~\cite{scheideler:shell} and in social networks~\cite{yu:survey,yu:sybilguard,mohaisen:improving,wei:sybildefender,Yu:2010:Sybillimit,alvisi2013sok,lesniewski-laas:whanau}. However, the information required---particularly social networks---may not always be available. 

An alternative defense is to use measurements of communication latency or wireless signal strength to verify the uniqueness of IDs~\cite{bazzi:distinct,1550961,liu:mason,Gil-RSS-15,demirbas:rssi}. However, these techniques are sensitive to measurement accuracy. 

For DHTs, an early result by Danezis et al.~\cite{danezis:sybil} gives a heuristic to limit the impact of bad IDs using bootstrapping information, but unfortunately provides no formal guarantees.  Results that employ resource burning are scarce.  The use of computational puzzles in decentralized systems is explored by Borisov~\cite{borisov:computational} and Tegeler and Fu~\cite{tegeler:sybil} as a means for identifying and excluding bad IDs from the system. Computational puzzles are also used by Rowaihy et al.~\cite{Rowaihy:2007} to throttle the rate of bad IDs added to a structured P2P system; however, this does not limit their number.  Arguably the best-known result is the SybilControl scheme by Li et al.~\cite{li:sybilcontrol}, which provides for a DHT construction that limits the number of bad IDs through the use of computational puzzles. Good IDs periodically challenge their neighbors under the Chord DHT topology~\cite{stoica_etal:chord,DBLP:journals/ton/StoicaMLKKDB03}, and blacklist those who do not respond with a solution in time. Experimental results indicate that this approach, in conjunction with limited data replication, allows for almost all searches to succeed. 

\subsection{Why \defID is Not Enough} The \defID problem (Section~\ref{sec:gendefid}) captures many of the challenges required for secure DHTs. However, current solutions to \defID depend heavily on a means to coordinate resource burning.  The main approach is to use a committee---a small set of IDs with a good majority---which issue resource-burning challenges. To apply results on \defID to DHTs requires decentralizing the functionality provided by the committee.

Additionally, while \defID always guarantees a minority of bad IDs, this is not enough.  In particular, to ensure reliable routing and protection from eclipse attacks, group-based approaches demand that all groups have a minority of bad IDs.  Fortunately, there are already clever techniques to spread the bad IDs uniformly across the groups.  Informally, when a new ID joins a group, some IDs in the group are evicted and resettled in random locations, and their replacements are selected uniformly at random~\cite{awerbuch:towards,awerbuch:towards2,awerbuch:random,guerraoui:highly}.  

Unfortunately, performing such shuffling for every joining ID,  even when there are no bad IDs in the system, incurs large bandwidth costs. A major open problem is to devise an algorithm that minimizes both bandwidth and resource-burning costs, as a function of adversarial spend rate.

\begin{figure}[h]
\centering
\begin{tcolorbox}[standard jigsaw, opacityback=0]
\begin{minipage}[h]{1.0\textwidth}

\begin{problem}\label{op:dht}
\noindent{\bf A Secure DHT in the Permissionless Setting}
\end{problem}
\noindent{\bf Model:}  The adversary has complete control over the scheduling of joins and departures for bad IDs and limited control for good IDs. There is no explicit assumption that the good IDs are in the majority at all times.\smallskip

\noindent{\bf Goal:} A DHT that enables secure and efficient routing between any two good IDs in the system. 
\end{minipage}
\end{tcolorbox}
\label{position-dht}
\vspace{-15pt}
\end{figure}

\subsection{The Permissionless DHT Problem}
Problem~\ref{op:dht} gives our formal problem in this domain.  It assumes that the adversary controls an $\alpha<1/3$ fraction of the burnable resource.  We now describe some ideas about how to solve it.

Recall from Section~\ref{sec:gendefid} that \defID imposes a cost of $O(J_G + \sqrt{TJ_G})$ on the good IDs. Informally, a plausible extension to this result is for each group in the DHT to act as a committee that runs an algorithm to solve \defID. In many group-based constructions, a good ID belongs to a number of groups that is logarithmic in the system size. Consequently, the algorithmic spend rate is likely to increase by a logarithmic factor.  This yields our conjectured bound of $\tilde{O}(\sqrt{T J_G} + J_G)$. Note that this aligns with Position 2 since costs to the good IDs are low when the adversary expends little effort (or does not attack at all), and grows slowly relative to the adversary's cost when a significant attack occurs. In the absence of a single committee that can track global information (such as the join rate of IDs), setting the hardness of challenges is tricky, and new ideas are needed to obtain the conjectured upper bound. 

Finally, while we have focused on DHTs, new defenses for them might generalize to providing security in permissionless settings for other structured P2P systems~\cite{fraigniaud:d2b,jagadish:baton,harvey:skipnet,aspnes:skip,awerbuch:hyperring,zatloukal_harvey:family,fiat_saia:censorship}. 


\section{Application-Layer DDoS Attacks}\label{sec:ddos}

A denial-of-service (DoS) attack prevents good IDs from accessing resources of a system. A distributed denial-of-service (DDoS) attack occurs when multiple  bad IDs carry out a coordinated DoS attack.  In an application-layer DDoS attacks,  an adversary attacks by issuing many requests for system resources, as opposed to say swamping the network bandwidth. Here, we discuss defenses against application-layer DDoS attacks based on resource burning.\medskip

\noindent{\bf Filtering Methods.} Many DDoS defenses rely on techniques for filtering out malicious traffic, including IP profiling \cite{malliga2008filtering,yu2009discriminating}; CAPTCHAs~\cite{von2003captcha,oikonomou2009modeling};  capability-based schemes~\cite{anderson2004preventing,yang2008tva}\footnote{Informally, this refers to a scheme where the source makes a ``capability'' request and, if approved by the receiver, will then obtain prioritized service from those routers along the path between the source and the receiver.}; and anomaly detection~\cite{hussain:framework}. An extensive survey of defenses can be found in~\cite{6489876}. Unfortunately, these techniques are imperfect, and an adversary may bypass them by issuing traffic that appears legitimate.  This has led to resource-burning defenses against DDoS attacks, which are sometimes referred to in the literature as currency-based or resource-based schemes~\cite{walfish2010ddos}. 

\smallskip

 \noindent \textbf{Resource-Burning Approaches.}  A number of proposed defenses require IDs to solve puzzles before their requests for service are honored~\cite{juels:dos,aura2000resistant,parno2007portcullis,kaiser:kapow}.  A challenging aspect of these proposals is the lack of a theoretically-backed method to tune the puzzle difficulty. To address this issue, Mankins et al.~\cite{mankins2001mitigating} propose a pricing mechanism to set the difficulty based on the service-request type; however, the pricing functions are set by the server {\it a priori}, and may fail as the incentives or capabilities of the attacker change over time. A dynamic strategy to determine puzzle difficulty is given by Wang and Reiter~\cite{wang:defending}. A client requesting service chooses the puzzle difficulty based on the effort it is willing to expend, while the server  prioritizes service according to the difficulty of the puzzles solved. However, this approach may starve IDs with limited resources, and requires the server to maintain state on the difficulty of the puzzles solved. Finally, Noureddine et al.~\cite{noureddinerevisiting} employ a game-theoretic model to pre-compute the difficulty of puzzles assuming all IDs (good and bad) are rational. 
 
An alternative resource---communication capacity---is consumed by the {\it speak-up} defense of Walfish et al.~\cite{walfish:ddos}. During an attack, it is common for bad IDs to bombard the server with requests, using much (or all) of the data rate available to the adversary. Speak-up encourages good IDs to respond in kind by increasing their respective request rates. A front-end server known as a ``thinner'' randomly drops requests in order to impose a manageable service load. If the aggregate capacity of the good IDs is comparable to that of the bad IDs, then this resource-burning scheme can allow good IDs to obtain a commensurate amount of service.



\subsection{The Application-Layer DDoS Problem} 

There are many similarities between the application-layer DDoS attack and the Sybil attack. The DDoS model is not purely permissionless, since the server is a trusted authority.  However, the attacks involve IDs whose distinctness cannot be ascertained, and where an adversary may create many bad IDs to facilitate attacks.  In this sense, the DDoS model is a hybrid of permissionless and permissioned systems.  Thus, it is not surprising that resource burning would be useful to defend against DDoS attacks. 

In this vein, we propose the open problem below. \vspace{-5pt}

\begin{figure}[H]
\centering
\begin{tcolorbox}[standard jigsaw, opacityback=0]
\begin{minipage}[h]{1.0\textwidth}

\begin{problem}\label{op:dos}
\noindent{\bf Application-Layer DDoS Attacks}
\end{problem}
\noindent{\bf Model:} There are $n$ good client IDs and a good server.  An adversary  controls an $\alpha$-fraction of the consumable resource, and can generate any number of bad client IDs.  Client IDs can request service from the server at any time.  The server must decide which requests to service based on its own limited resources.

\smallskip
\noindent{\bf Goal:} The good clients obtain a $1-O(\alpha)$ fraction of the service provided by the server.
\end{minipage}
\end{tcolorbox}
\label{insider-DDoS}
\vspace{-5pt}
\end{figure}

Problem~\ref{op:dos} shares much in common with \defID (Section~\ref{sec:gendefid}). Requests from client IDs correspond to join events; satisfying requests corresponds to departures. Here, $\alpha$ need not be bounded, since we are not making a correctness guarantee analogous to maintaining a good majority in \defID. Rather, our new requirement concerns performance: good IDs receive a $1-O(\alpha)$ fraction of service. In this sense, Problem~\ref{op:dht} seems strictly easier than \defID.

However, a new difficulty is heterogeneity: requests may differ in the amount of effort required to service them. Thus, enforcing a bound on the fraction of bad requests serviced does not ensure that the goal of Problem~\ref{op:dos} will be met. In light of this issue, it may be helpful to consider a weighted version of \defID, and whether existing solutions can be extended to this more general setting.  While we are optimistic that for large $T$, $o(T)$ is possible for Problem~\ref{op:dos}, a tight upper bound is an interesting direction for future work. 


\section{Review Spam}\label{sec:spam}


Online user-generated reviews play an important role in influencing the purchasing decisions of consumers. These systems are subject to manipulation where an adversary employs multiple accounts to create fake reviews that falsely promote or disparage a product~\cite{forbes:spam}; this malicious behavior is often referred to as \defn{review spam}, but also goes by other labels such as {\it astroturfing}~\cite{guardian:astroturfing} and {\it opinion spam}~\cite{MATHEWSHUNT20153}.

Review spam threatens  online retailers---such as Amazon or Walmart~\cite{cbs:spam,forbes:spam}---and merchants who depend on income from online sales. While online review systems typically have some form of admission control, such as requiring credentials for the creation of an account, this can be bypassed. For example, an attacker can hire users that possess a sufficient online presence in order to engage in review spam~\cite{cracked:spam,hooi:fraudar}, and social-media credentials can be automatically generated~\cite{guardian:astroturfing}; examples of these attacks  are described in~\cite{malbon2013taking,MATHEWSHUNT20153}. 


In response to this threat, the research community has proposed various strategies for detecting fraudulent reviews; these employ a range of techniques including machine learning~\cite{chau2006detecting,jindal2008opinion}, anomaly detection~\cite{savage:detection,sihong:review,xie:review}, linguistic evaluation~\cite{karami2015online,shebuti:collective}, graph analysis~\cite{hooi:fraudar,beutel:copycatch,akoglu2013opinion}, and many others.  A comprehensive overview of these techniques is given in~\cite{ma:detecting,WU2020113280,HEYDARI20153634}.

Progress in this area offers the ability to classify a review as either spam or legitimate, with some small error probability; for example, the work in~\cite{ott:finding} achieve an accuracy of almost $90\%$.  This classification functionality is a promising ingredient for designing more general tools for mitigating review spam.


\subsection{The Review Spam Problem} 


The problem of review spam largely aligns with our general model in Section~\ref{sec:model-main}. While online systems often require some credentials for creating an account, this admission control can be circumvented, and the system is effectively permissionless. However, the review spam model has some novel features. IDs join the system, but they may never formally depart. Even IDs that are regularly in use may have periods where the corresponding user is offline.  Thus, any attempt to simultaneously challenge all IDs, in order to reveal some as bad, will fail.


On the positive side, as noted above, machine learning can now help.  In particular, we may assume a classifier that correctly classifies reviews as spam or not with some fixed probability of error.  Over a sufficiently large number of reviews, this classifier can be used to obtain a good approximation of the current fraction of spam reviews, and this information can be used to set the amount of resource burning required to post a review. Our conjecture of $O(T^{2/3} + P_G)$ in Table~\ref{table:overview} follows from a preliminary analysis that leverages a classifier in this way. Informally, we increase the cost for posting a review when a significant attack is ongoing---that is, many reviews are diagnosed as spam by the classifier. Otherwise, we reset the cost to the lowest level. 

We formalize the challenge of review spam as Problem~\ref{op:spam}. 

\begin{figure}[h]
\vspace{-5pt}
\centering
\begin{tcolorbox}[standard jigsaw, opacityback=0]
\begin{minipage}[h]{1.0\textwidth}

\begin{problem}\label{op:spam}
\noindent{\bf Review Spam}
\end{problem}
\noindent{\bf Model:} IDs post reviews online. A classifier labels each post as legitimate or as spam, with some fixed error probability. Each spam post has unit cost, reflecting its negative impact on system usability. The algorithm can also set an arbitrary resource-burning cost for each new post, based on the classification of past posts.

\smallskip

\noindent{\bf Goal:} Minimize costs due to spam posts plus resource-burning costs incurred from legitimate posts.  
\end{minipage}
\end{tcolorbox}
\label{position-spam}
\vspace{-15pt}
\end{figure}


\section{Conclusion}

In this paper, we surveyed the literature on resource burning and established it as critical a tool for securing permissionless systems.  We described results from four domains: blockchains, DHTs, application-layer distributed DDoS attacks, and review spam. We noted shared security vulnerabilities in both permissionless and hybrid systems, and how resource burning is well-suited for addressing common threats. 

We observed that resource burning costs are prohibitively high for most current systems. Thus, a high-priority area for theoretical research is the design of resource-burning defenses that reduce these costs.  In particular, whenever possible, good IDs should spend at a rate which is asymptotically less than the adversary when the system is under attack.  To encourage research efforts, we defined several open problems, along with conjectured upper bounds for these problems.\medskip

\noindent{\bf Acknowledgements.} We are grateful to the organizers of SIROCCO 2020 for inviting this paper, and we thank Valerie King for  helpful feedback on our manuscript.

\end{document}